\documentclass[aps,pre,reprint,showpacs,floatfix,superscriptaddress]{revtex4-1}
\usepackage[colorlinks,linkcolor=red,anchorcolor=red,citecolor=blue,urlcolor=blue]{hyperref}
\usepackage{amsmath,txfonts}
\usepackage{graphicx}% Include figure files
\usepackage{dcolumn,enumitem}% Align table columns on decimal point
\usepackage{float}
\allowdisplaybreaks[3]

\begin{document}

\title{Modularity-based selection of the number of slices in temporal network clustering}

\author{Patrik Seiron}
\affiliation{Department of Information Technology, Uppsala University, Sweden}
\author{Axel Lindegren}
\affiliation{Department of Information Technology, Uppsala University, Sweden}
\author{Matteo Magnani}
\affiliation{Department of Information Technology, Uppsala University, Sweden}
\author{Christian Rohner}
\affiliation{Department of Information Technology, Uppsala University, Sweden}
\author{Tsuyoshi Murata}
\affiliation{Department of Computer Science, Tokyo Institute of Technology, Japan}
\author{Petter Holme}
\affiliation{Department of Computer Science, Aalto University, Finland}
\affiliation{Center for Computational Social Science, Kobe University, Japan}

\begin{abstract}A popular way to cluster a temporal network is to transform it into a sequence of networks, also called slices, where each slice corresponds to a time interval and contains the vertices and edges existing in that interval. A reason to perform this transformation is that after a network has been sliced, existing algorithms designed to find clusters in multilayer networks can be used. However, to use this approach, we need to know how many slices to generate. This chapter discusses how to select the number of slices when generalized modularity is used to identify the clusters.\end{abstract}

\maketitle

\section{Introduction}

Clustering is one of the most studied network analysis tasks, with an ever-growing number of articles proposing new algorithms and approaches \cite{fortunato_community_2010,coscia_classification_2011,bothorel_clustering_2015}. The absence of a unique definition of a cluster can partly explain the large number of available clustering algorithms existing. While a cluster in a simple network is generally understood as a set of vertices that are well connected to each other and less well connected to other vertices, different algorithms use different specific definitions of cluster and clustering (that is, the set of all clusters). When we add more information to vertices and edges, e.g., the fact that edges only exist at specific times, defining what constitutes a cluster or a clustering becomes even more challenging.

The availability of temporal information about the existence of edges has two main consequences on the clustering task. First, clustering algorithms not considering the temporal information may miss clusters only existing at specific times (because they are hidden by edges active at other times) or identify clusters that do not exist at any specific time. Second, new types of clusters can be defined, for example, clusters recurring at regular times or clusters growing, shrinking, merging, and splitting \cite{palla2007groupevolution}. For these reasons, different extensions of clustering methods for temporal networks have been proposed.

\begin{figure}
    \includegraphics[height=2cm]{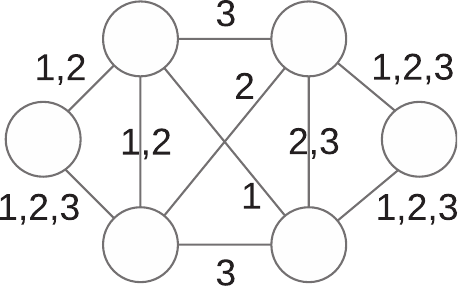}
    \caption{A temporal network with three timestamps}
    \label{fig:ex1}
  \end{figure}

\begin{figure}
    \includegraphics[width=\columnwidth]{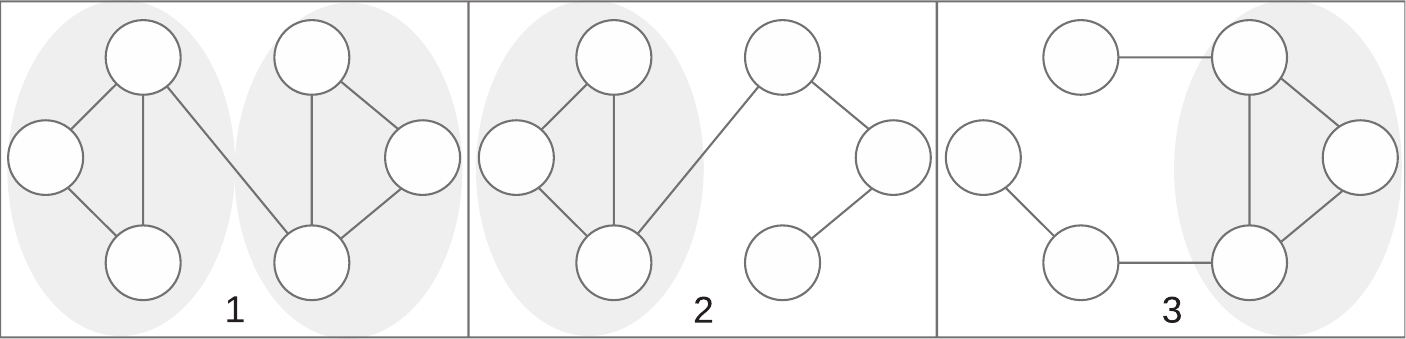}
    \caption{Three temporal slices, one for each timestamp, where distinct clusters can be identified}
    \label{fig:ex2}
  \end{figure}

In this chapter, we focus on a common approach to temporal network clustering, consisting of two steps: first, the temporal network is sliced into a sequence of static networks, also called slices. This sequence of networks is a specific type of multilayer network \cite{kivela_multilayer_2014,boccaletti_structure_2014,dickison_multilayer_2016}. Then a clustering algorithm for multilayer networks is used to discover clusters. An example of this approach is shown in Figure \ref{fig:ex1} and \ref{fig:ex2}.

While this approach allows reusing the many clustering algorithms defined for multilayer networks, it relies on the ability to choose a number of slices leading to good clustering. This number can sometimes be decided based on domain knowledge. Still, for this approach to be usable in general, it is important to have ways to discover a good number of slices directly from the data. This is fundamental in the absence of domain knowledge, but also useful to check if the number suggested by a domain expert is compatible with the data. Therefore, the research question we address in this chapter is: given a temporal network and a multi-slice network clustering algorithm, how can we find a number of slices for which well-defined clusters emerge?

The answer to this question depends on the algorithm used to cluster the network after slicing. In this paper, we focus on one of the most used multilayer network clustering algorithms: generalized Louvain \cite{mucha_community_2010}. Multi-slice modularity, the objective function used by the algorithm, considers modularity in each layer and also increases when the same vertex is included in the same cluster in different layers. 

If our objective is to find the number of slices leading to the best clustering, having an objective function (in this case, multi-slice modularity) we might be tempted to run the generalized Louvain optimization algorithm for different numbers of slices and pick the result with the highest modularity. Given the same network (or the same sliced sequence of networks), we can compare the modularity of different clusterings to identify the best. In particular, if two clusterings of the same network sliced into the same number of slices have a different modularity, it is often assumed that the clustering with a higher modularity is a better clustering.
Unfortunately, in general, we cannot use modularity to compare clusterings of the same network sliced differently. If we split the same network using two different numbers of slices, a clustering of the former with higher modularity is not necessarily better than a clustering of the latter with lower modularity.

\begin{figure}[ht]
\centering
\includegraphics[width=\linewidth]{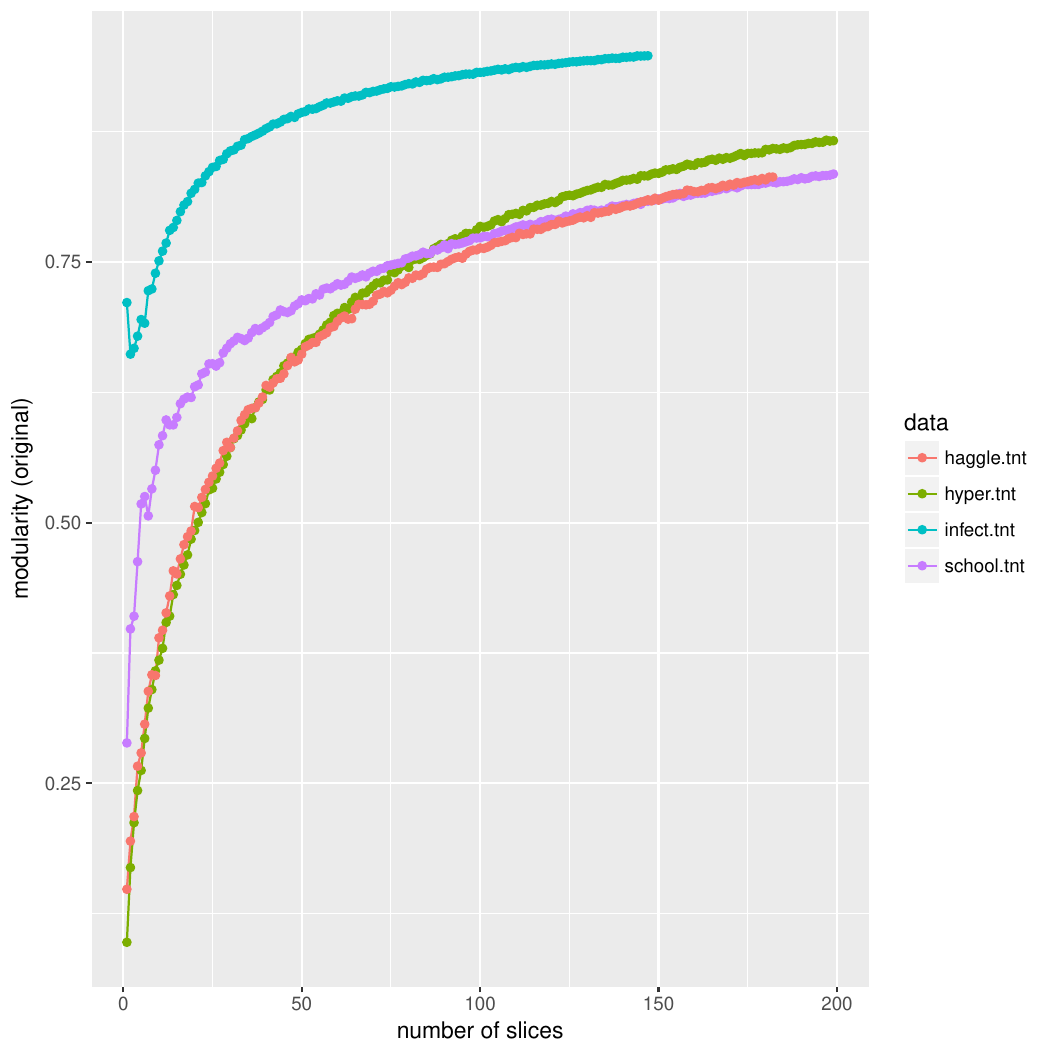}
\caption{Modularity of the partitions returned by the generalized Louvain algorithm varying the number of slices for four real temporal networks.}
\label{fig:stream}
\end{figure}

As an example, Figure~\ref{fig:stream} shows the modularity of the clusterings discovered by the generalized Louvain algorithm on four real temporal networks varying the number of slices. The four networks represent contacts between people measured by carried wireless devices (haggle \cite{chaintreau_impact_2007}), face-to-face interactions (hyper, infect \cite{isella_whats_2011}), and friendships between boys in a small high school in Illinois (school \cite{coleman_introduction_1964}). We can see that the more slices we have, the higher the modularity we get from the algorithm. This suggests that increasing modularity values for different numbers of slices is not necessarily an indication of a better clustering but can be a by-product of the changing size of the input networks.

To address this problem, we use the following hypothesis. Multi-slice modularity has two components: one that increases with better clusterings and one that increases just because the data size increases, e.g., if we duplicate a slice, the same cluster extended across two slices will contain additional inter-slice edges. If this hypothesis is correct, then we can try to isolate the first component in the modularity value and use it to compare clusterings computed using different numbers of slices.

In this chapter, we show the dependency between the number of slices and multi-slice modularity, both analytically and experimentally. We also use an edge reshuffling algorithm to separate the effect of the number of slices, resulting in a first corrected multi-slice modularity measure. We then experimentally validate the corrected multi-slice modularity on synthetic networks where the best number of slices is known in advance, and we identify clusters in different types of real temporal networks using the proposed approach. We conclude by discussing the limitations of the reshuffling-based method and, more in general, of the modularity-based approach.

\section{Related work} 

For a detailed overview of clustering methods in temporal networks, we refer the reader to the dedicated chapter in this book.

Clustering algorithms for temporal networks approach the temporal evolution of networks in different ways. 
% segmentation
Algorithms building on segmentation seek time slices containing well-defined community structures (quality) yet being similar to neighboring slices (stability). For example, Aynaud, et al.~\cite{aynaud2011hierarchical} builds a hierarchical time segmentation by extracting interesting time windows based on structural changes and identifying a unique decomposition for the time windows. He, et al.~\cite{he2017stepwisedetection} uses the so-called Moore’s Visualization Method, which implies a certain overlap of time slices. Simpler approaches segment the temporal network in equal time slices or slices with similar edge density. In this chapter, we abstract from the segmentation approach by using equal time slices and focus on how many slices should be used.
%% MM simplified a little

After a temporal network has been converted into a multilayer network, several approaches can be used to discover clusters. These include, among others, algorithms based on generative models \cite{de_bacco_community_2017}, multilayer cliques \cite{DBLP:conf/socinfo/TehraniM18}, walks \cite{Boutemine2017}, information theory \cite{de_domenico_identifying_2015}, aggregation of single-layer clusterings \cite{berlingerio_abacus_2013,Tagarelli2017}, and modularity \cite{mucha_community_2010}. An overview of existing approaches is available in \cite{magnani_community_2021}. Different algorithms using different definitions of cluster require different measures to evaluate the goodness of a slicing, and the best number of slices does not need to be the same for all approaches. In this chapter, we focus on modularity-based clustering.

% evolution tracking
Other algorithms avoid segmentation and instead track evolution in time by observing events (i.e., birth, death, growth, contraction, merge, split, continue, resurgence)~\cite{rossetti2017tiles}. Palla et al.~\cite{palla2007groupevolution} builds on clique percolation and observe that small communities, in general, have static, time-independent membership, while large communities are dynamic. Random walk-based approaches~\cite{rosvall2008maps_of_random_walks} and stochastic block modeling~\cite{matias2017sbm} are other alternatives to modularity-based clustering. 

\section{Method} 

To evaluate the quality of a clustering we use the concept of modularity, describing the fraction of the edges within communities minus the expected fraction if edges were distributed at random, preserving degree distribution. 
%In the most common definitions of modularity, the randomization of the edges is done using a configuration model to preserve the node degree of each vertex. Formally, the value of the modularity is derived as
For simple networks, and without considering the so-called resolution parameter that was introduced at a later time, modularity is defined as follows:
\begin{equation}
\frac{1}{2m}
\sum_{i,j} \left(A_{ij} - \frac{k_{i}k_{j}}{2m}\right)\delta(\gamma_{i}\gamma_{j})
\label{eq:modularity}
\end{equation}
where $A$ is the adjacency matrix, $k_{i}$ is the degree of vertex $i$, $\gamma_{i}$ is the community id of vertex $i$, $\delta(\gamma_{i}\gamma_{j}) = 1$ if $\gamma_{i} = \gamma_{j}$ and 0 otherwise, and $m$ is the number of edges in the network.

In multi-slice networks, modularity is defined as~\cite{mucha_community_2010}:
\begin{equation}
 \frac{1}{2 \mu} \sum_{ijsr} \left\{ \left( A_{ijs} - \frac{k_{is} k_{js}}{2m_s} \right) \delta(s,r) + c_{jsr} \delta(i,j)\right\} \delta(\gamma_{i,s},\gamma_{j,r})
\label{eq:multi-slice modularity}
\end{equation}
where $i$ and $j$ indicate vertices and $s$ and $r$ slices. This formula is a combination of the modularity in each slice plus a contribution $c_{jsr}$ for vertices included in the same community in two slices. In this chapter, we consider the case where $c_{jsr} = 1 \textrm{ iff } r = s + 1$, that is, the two slices are consecutive. $\mu$ is the number of all (intra-slice) edges plus the sum of all $c_{jsr}$.

The basic approach we use to separate the effect of the presence of clusters from the effect of just increasing the number of slices is based on an edge reshuffling process that destroys the clusters in the network without affecting the degree distribution. For each number of slices, the Louvain algorithm is run both on the original data and on the reshuffled data where the clusters have been destroyed. The modularity on the dataset without clusters indicates the effect of the number of slices on modularity. Here we use the difference between the two to estimate the part of modularity due to the presence of clusters. We call this difference \emph{corrected modularity}.

%\subsection{Edge randomization}

%Also the second method to find a good number of slices is based on modularity.
In particular, we run community detection multiple times and take the maximum value of modularity to account for the non-deterministic character of the generalized Louvain algorithm. To remove the effect of the number of slices, we use the edge swapping randomization model \cite{karsai2011smallbutslow, rrm} that selects two edges $(i, j)$ and $(u, v)$ at random and swaps two randomly selected ends of the two edges. Some practical decisions have to be made to perform the randomization. First, we have to avoid swaps producing existing edges so that the total number of edges does not change. We must also do the shuffling slice by slice to preserve the intra-slice degree distributions --- that is, in this context, we do not use the exact same reshuffling as described in \cite{karsai2011smallbutslow, rrm}. It should also be noted that using this process, we cannot break a single clique contained in a single slice. 

In summary, the method can be described as follows:
\begin{enumerate}
	\item For $i$ from $1$ to $n$:
 \begin{enumerate}
	\item Slice the temporal network into $i$ slices.
	\item Run modularity-based multilayer community detection multiple times and compute modularity. We call $m_o(i)$ the maximum modularity found for $i$ slices. 
	\item Apply edge randomization in each slice. 
	\item Run community detection multiple times and compute modularity. We call $m_r(i)$ the maximum modularity found for $i$ slices after randomization.
	\item Compute the corrected modularity $m_n(i) = m_o(i) - m_r(i)$.
\end{enumerate} 
\item Return $i$ maximizing $m_n(i)$.
\end{enumerate}

%This approach is based on a null model that destroys the community structure of the network. The community structure is destroyed by edge randomization that renders the network into a random network, for which by definition, any result of a community detection is random. The data size can therefore be considered the only contributor to the value of the modularity. 

\section{Results}

In this section, we present three main results. First, we support our assumption that part of the growth in modularity when the number of slices increases is a direct consequence of the increased number of slices and not necessarily of the presence of better clusters. Then, we use synthetic datasets where the best number of slices is known in advance to test if our approach can correctly identify these values. Finally, we execute our method on real networks for which we do not know the best number of slices.

\subsection{Expected modularity increment in sequentially duplicated networks} 

In this section, we analyze the behavior of modularity when we start from a single slice with clear clusters and add additional identical slices. In this dataset, the assignment of vertices ($i, j, \dots$) into clusters does not change with the number of slices: the clusters are replicated on all slices.

We first do this analytically. We start from a single slice, where modularity is computed as in Equation \ref{eq:modularity}. For convenience, let us call $\overline{A} = \sum_{i,j} (A_{ij})\delta(\gamma_{i}\gamma_{j})$ and $\overline{K} = \sum_{i,j} (\frac{k_{i}k_{j}}{2m})\delta(\gamma_{i}\gamma_{j})$. With this substitution, we can write the modularity of one slice as: 
\begin{equation}
\frac{\overline{A} - \overline{K}}{2m}
\end{equation}

If we duplicate the network, that is, we add one slice identical to the original network and replicate the same cluster assignments on the two slices, the expected modularity includes the modularity on the two slices (each equal to the modularity above) plus one interlayer link for each vertex (as by definition each vertex belongs to the same cluster in both slices):
\[\frac{(\overline{A} - \overline{K}) + (\overline{A} - \overline{K}) + 2a}{2(a + m + m)}\]
Generalizing to $S$ slices, we obtain:
\[\frac{S(\overline{A} - \overline{K}) + 2a(S-1)}{2(a(S-1) + Sm)}\]

\begin{figure}[ht]
\centering
    \includegraphics[width=\columnwidth]{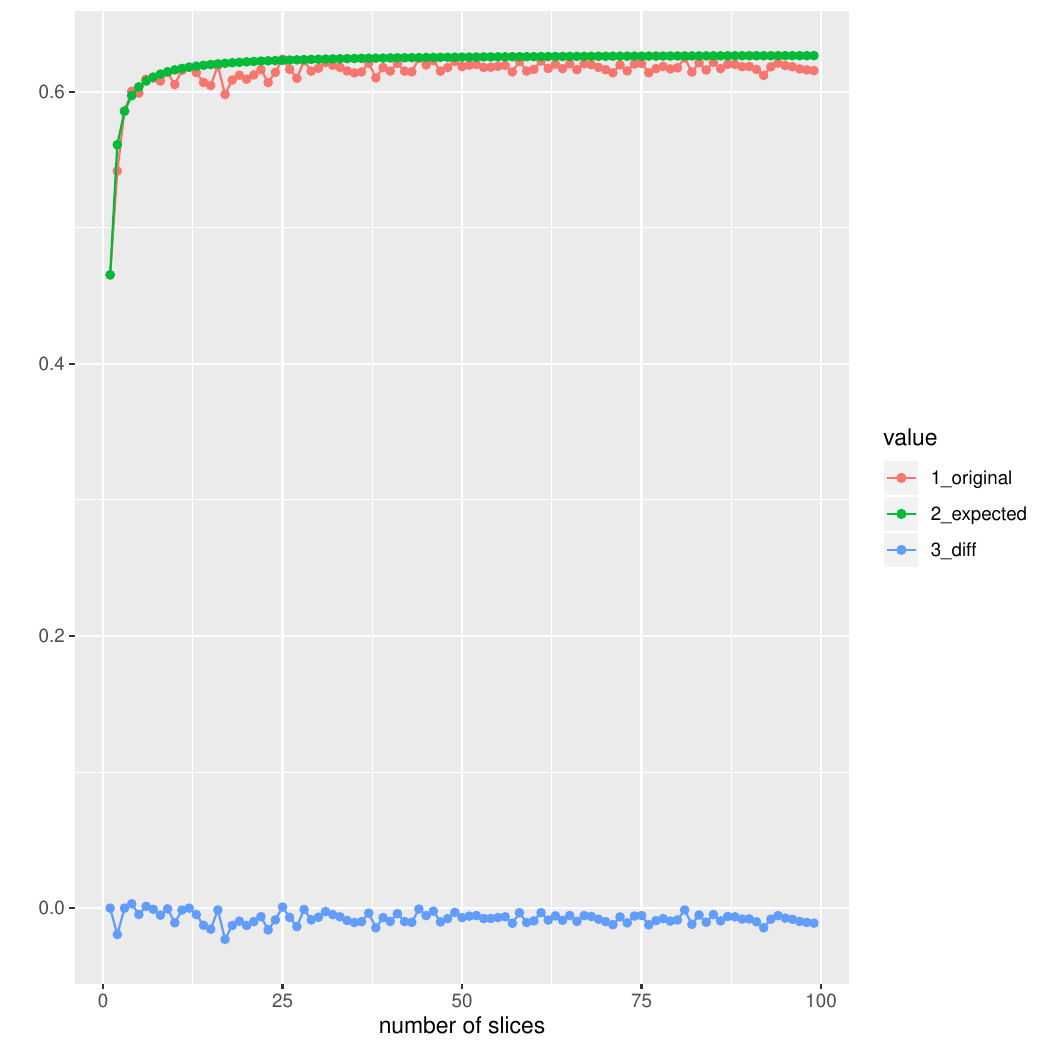}
    \caption{Expected modularity of a replicated network: the value increases with the number of slices, following a predictable pattern.}
    \label{fig:zac}
\end{figure}

This result is tested in Figure \ref{fig:zac}, where we show how the theoretical model approximates well the empirical one.

\subsection{Synthetic data validation} 

\subsubsection{Hidden cliques} 

\begin{figure*}[ht]
    \includegraphics[width=0.8\textwidth]{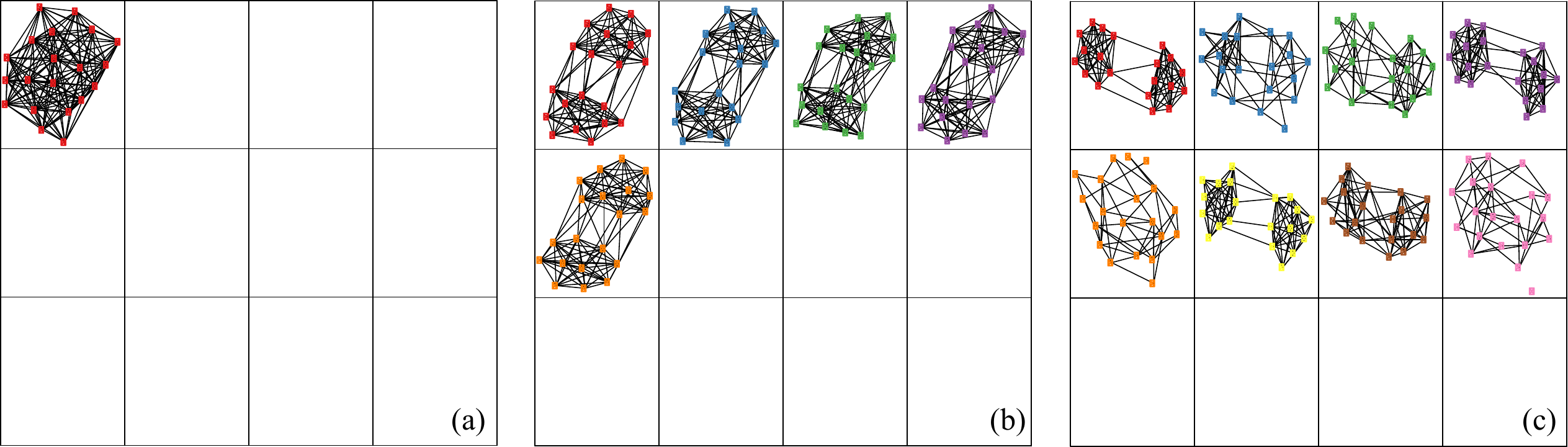}
\caption{A temporal network with two cliques separated by random noise is repeated five times: For a single slice (a), the aggregated noise hides the cliques. Five slices (b) reveal the cliques as the number of slices separates the five repetitions of the network. The cliques disappear for an increasing number of slices (c) because they are spread across several less dense slices.}
\label{fig:2c}
\end{figure*}

Our first synthetic data consists of two cliques separated by random noise (20\% density), with this pattern repeated five times. The network is shown in Fig~\ref{fig:2c}(a), split into different numbers of slices. When we only have one slice, the combination of the noise present throughout the existence of the network hides the clusters. When we use five slices (Figure \ref{fig:2c}(b)), the cliques are easily visible in all slices. In time, the cliques disappear because they are spread across several less dense slices.

\begin{figure*}[ht]
    \includegraphics[width=0.8\textwidth]{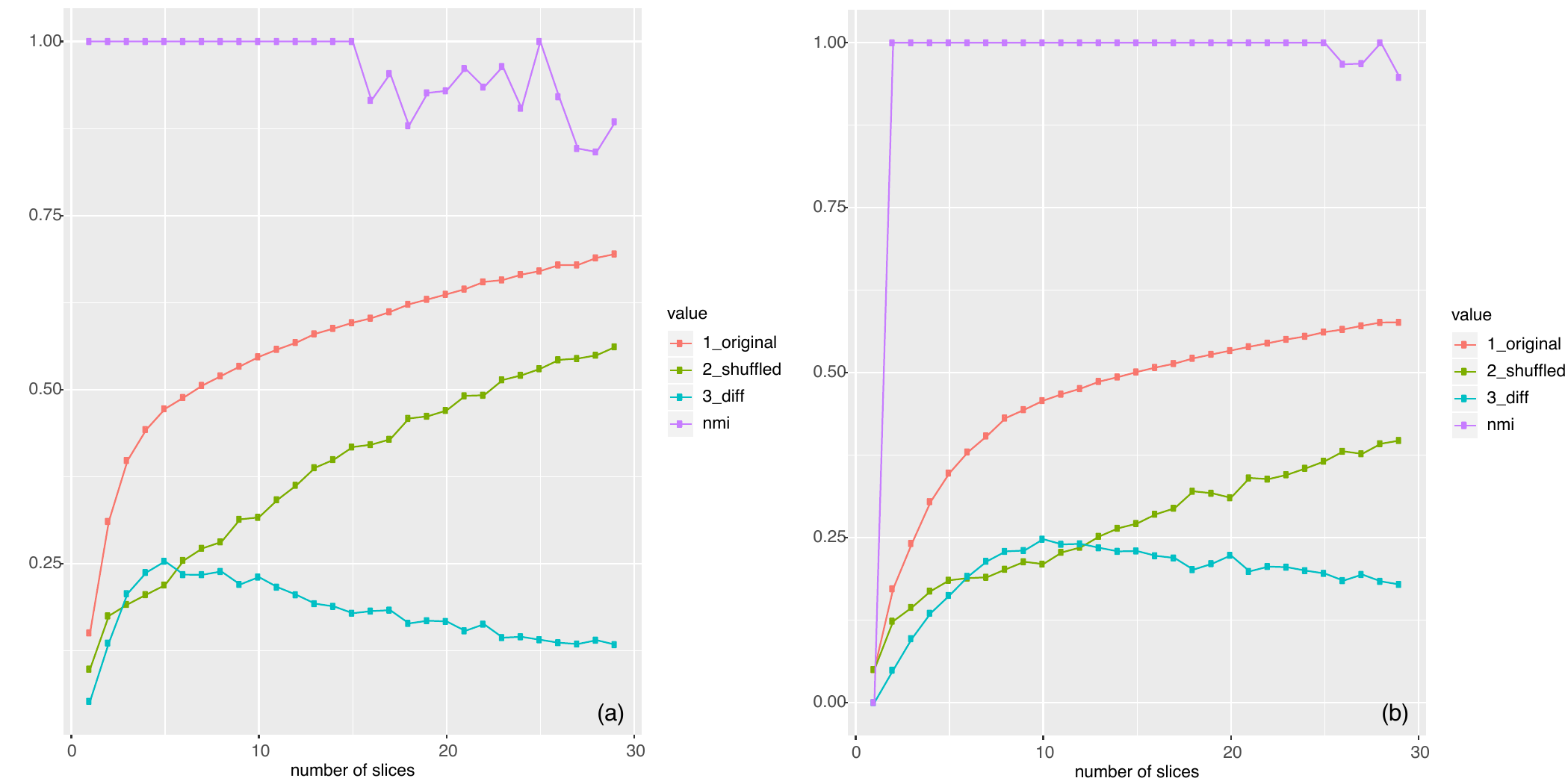}
\caption{A temporal network with two cliques separated by random noise is repeated five and ten times, respectively: The figures show the modularity of the original network (red), the shuffled network (green), the corrected modularity (blue), as well as the normalized mutual information NMI (purple), for five (a) and ten (b) repetitions of the experiment The peaks of the corrected modularity are at five and ten, respectively.}
\label{fig:2c_}
\end{figure*}

With this dataset, we know that the best clusters appear when we have five slices. Figure \ref{fig:2c_}(a) shows the original modularity, the randomized modularity, and our corrected modularity. While the first two increase when the number of slices increases, the corrected modularity peaks at five slices. Figure~\ref{fig:2c_}(b) shows the result for the same experiment but with ten repetitions of the clique-noise pattern instead of five. The method correctly finds a peak at ten slices.

Figure \ref{fig:2c_}(b) also shows the normalized mutual information (NMI) between the ground truth clusters and the clusters found by the algorithm, for different numbers of slices. A higher value of NMI corresponds to more similar clusterings. We notice how the number of slices identified by our approach corresponds to the highest NMI. However, the generalized Louvain algorithm would still be able to reach the same NMI with other numbers of slices (up to 15, respectively 25).

\begin{figure*}[ht]
    \includegraphics[width=0.8\textwidth]{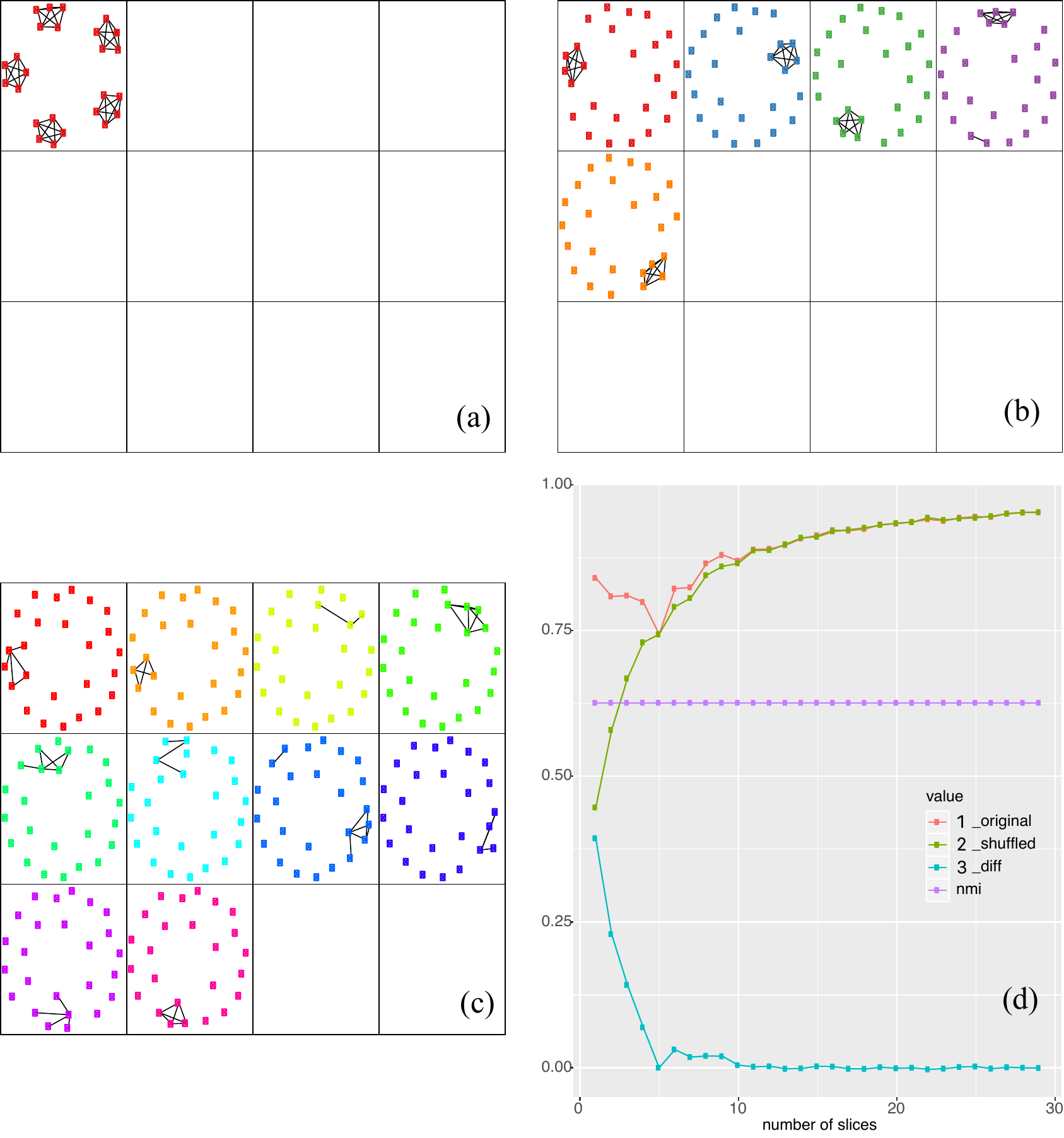}
\caption{Five cliques are active at different times: The five cliques are easily identified with one slice (a), where also the corrected modularity has its maximum. The normalized mutual information reveals, however, that the clusters are correctly identified independent of the number of slices---five in (b) and ten in (c). Panel (d) shows the modularity.}
\label{fig:ct}
\end{figure*}

Figure~\ref{fig:ct} shows a different type of synthetic temporal network where five groups of vertices are active at different times. An example of this type of behavior can be a museum, where some organized groups enter the exhibition at different times and go through it together, being active for the whole duration of their visit and then disappearing from the data.

In this case, the best number of slices is the one where all the groups are collected together, that is, only one slice. This corresponds to the highest value of corrected modularity (Figure~\ref{fig:ct}(d)). Notice, however, that the generalized Louvain algorithm would identify the same clusters independently of the number of slices.

\subsection{Real data} 

\begin{figure*}[ht]
    \includegraphics[width=0.8\textwidth]{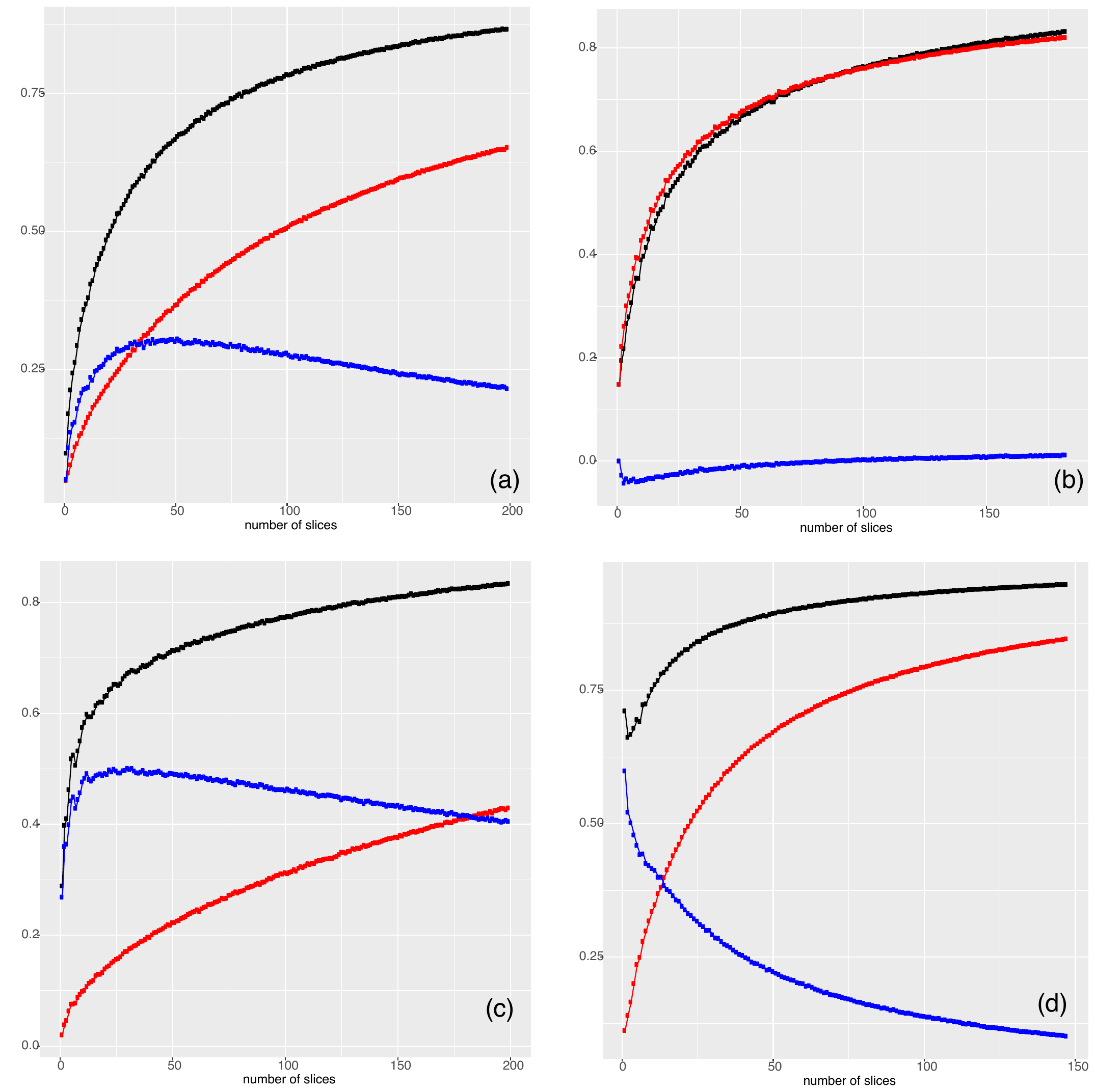}
\caption{Examples from real-world data sets. Hypertext (a), Haggle (b), School (c), Infect (d). original modularity (black) model reshuffling (red), and corrected modularity (blue).}
\label{fig:method2}
\end{figure*}

Figure~\ref{fig:method2} shows the original modularity computed by the generalized Louvain algorithm (black), the modularity of the randomized network (red), and the corrected modularity (blue) for the four real datasets in Figure~\ref{fig:stream}. 
The corrected modularity follows similar trends for the Hypertext, Haggle, and School datasets, with a maximum reached after a few slices have been obtained, while for the Infect dataset, we see a maximum for the original, unsliced data.

\section{Discussion}

We have presented an approach to select the number of slices to discover communities in temporal networks using generalized modularity. Based on the observation that the value of generalized modularity tends to increase when we increase the number of slices, even when no additional information is generated, the proposed approach corrects the modularity. In this chapter, we use edge reshuffling to perform the correction.

While the proposed method looks for the highest value of corrected modularity to select the best number of slices, this does not mean that lower values of corrected modularity would necessarily correspond to worse clusterings. In fact, even when the proposed method identifies the number of slices corresponding to the best clustering, our experiments show that there is often a range of values where the communities are clear enough to be identified by the clustering algorithm.

This chapter does not focus on efficient computation. In the experiments, we use a brute-force search to find the number of slices maximizing the corrected modularity, trying all values from 1 to an arbitrary number. While this works for small or medium networks, recomputing generalized Louvain hundreds of times may not be feasible for larger networks, so a smarter exploration of the solution space would be necessary.

\begin{figure}[ht]
\centering
\includegraphics[width=\columnwidth]{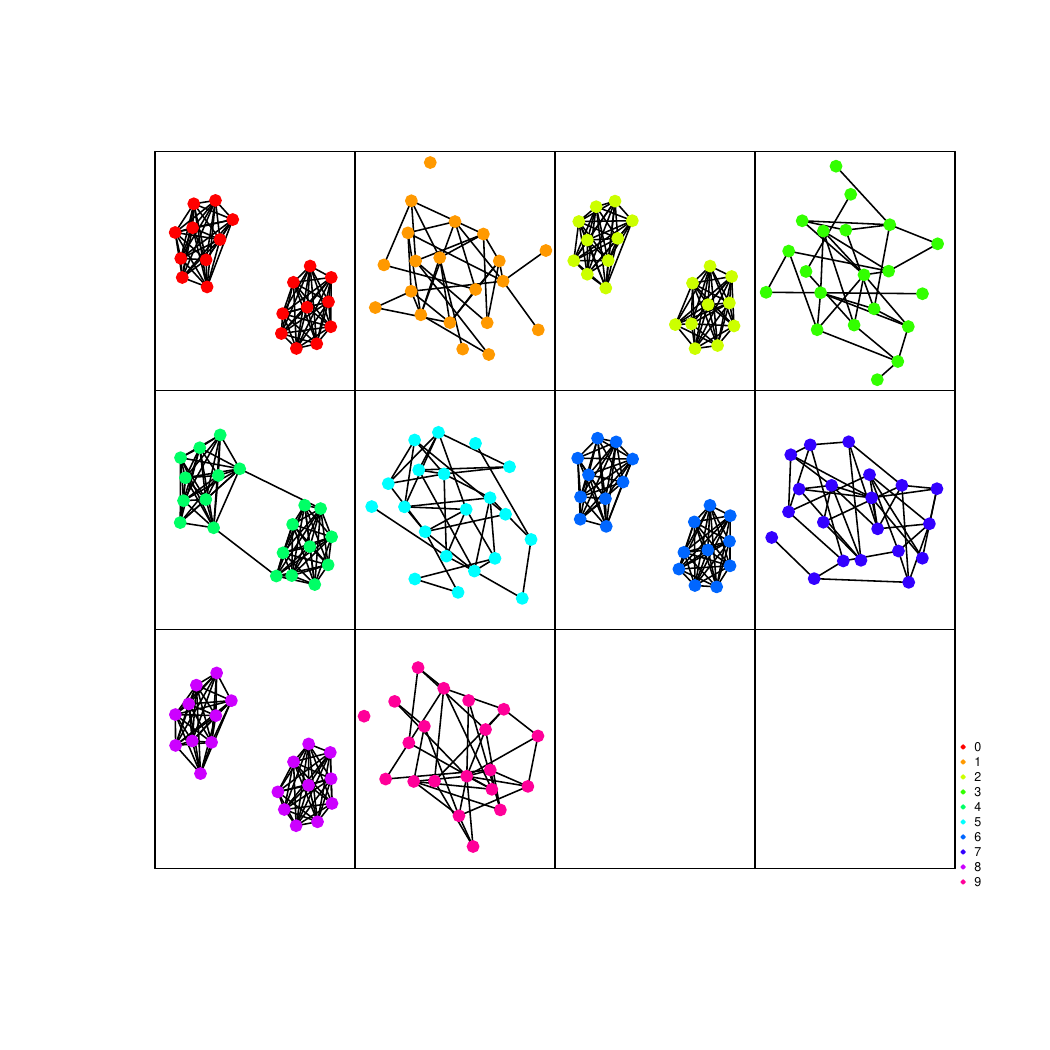}
\caption{An example of recurrent communities that are not captured well by multi-slice modularity.}
\label{fig:recurrent}
\end{figure}

While this chapter focuses on modularity, there are intrinsic limitations of modularity that should be considered when it is used to identify communities in temporal networks.
A first issue is that generalized Louvain tends to identify pillar clusters so that a vertex that belongs to the same cluster in many layers may get clustered with the same nodes also in layers where they are not well connected. More generally, in a temporal network, we may expect some vertices to belong to some clusters only at some times, while generalized Louvain would force all vertices to belong to a cluster in each slice. Recurrent clusters (appearing and disappearing) are also not supported well by the ordered version of generalized Louvain. Figure \ref{fig:recurrent} shows a slicing where clear communities are visible, but that does not correspond to a maximum value of corrected modularity. This is expected, as modularity tries to cluster all the vertices and looks for inter-layer consistency.

\begin{figure}[ht]
\centering
\includegraphics[width=\columnwidth]{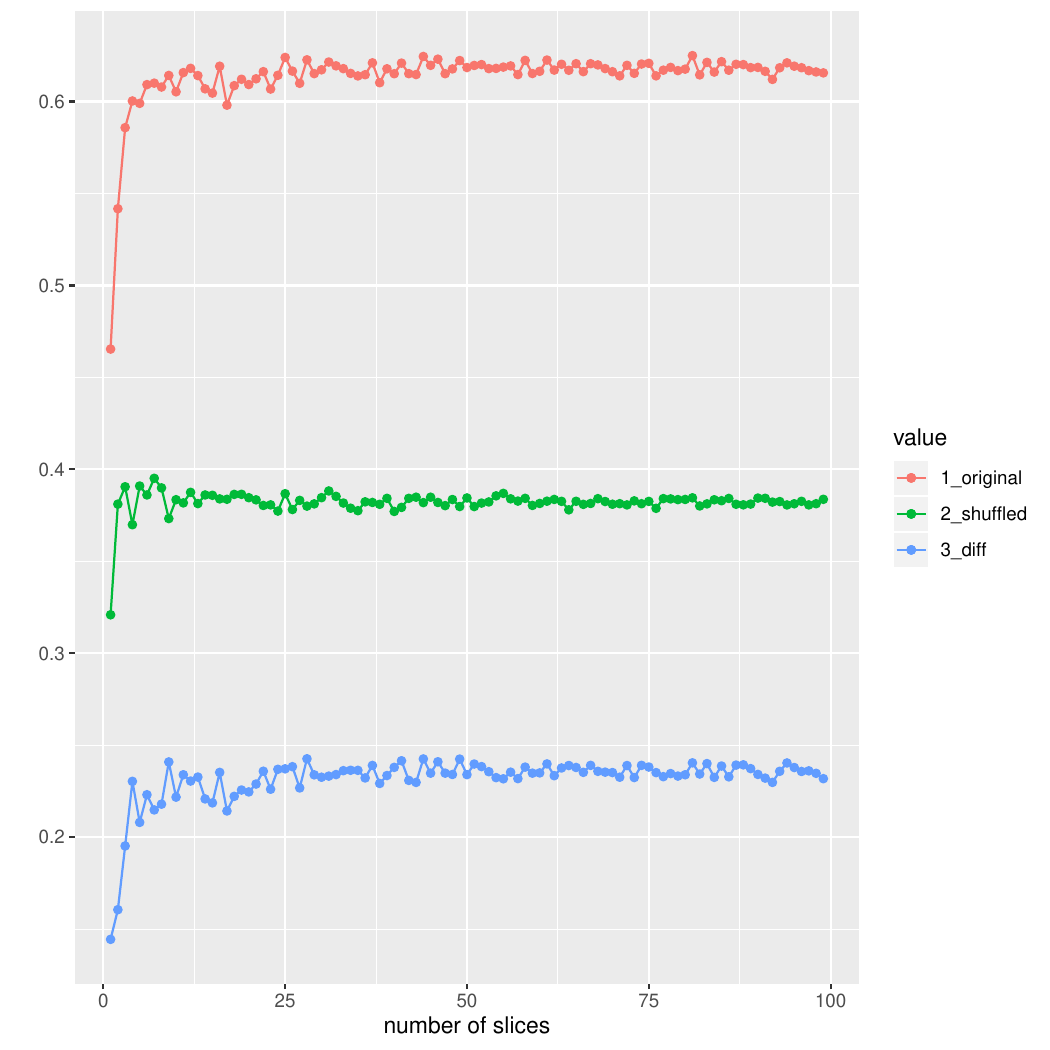}
\caption{The same network is replicated on all the slices: A scenario where the corrected modularity is expected to be constant but shows unexpected behavior for small numbers of slices.}
\label{fig:zac:shuffling}
\end{figure}

As a final consideration, in this chapter, we assume that the two modularity components, one that increases with better clusterings and one that increases just because the data size increase, are additive. The assumption seems to hold for identifying clearly planted clusters in synthetic data but also gives unstable results for low numbers of slices in a case where we repeat the same network in every slice (see Figure~\ref{fig:zac:shuffling}); in this case, we could expect corrected modularity not to show any difference between different numbers of slices.

\begin{acknowledgments}
This work has been partly funded by eSSENCE, an e-Science collaboration funded as a strategic research area of Sweden, by STINT initiation grant IB2017-6990 ``Mining temporal networks at multiple time scales'', and by EU CEF grant number 2394203 (NORDIS -- NORdic observatory for digital media and information DISorder). P.H. was supported by JSPS KAKENHI Grant Number JP 21H04595.
\end{acknowledgments}

\bibliographystyle{abbrv}
\bibliography{sample}

\end{document}